\documentstyle[aps,prd]{revtex}
\tighten

\def \D {\mbox{D}}
\def \c {\mbox{curl}\,}
\def \div {\mbox{div}\,}
\def \ep {\varepsilon}
\def \ts {\textstyle}
\def \rd {\displaystyle{\cdot}}

\begin{document}

\title{Nonperturbative gravito-magnetic fields}

\author{
Carlos F. Sopuerta\dag\S\,,
Roy Maartens\ddag\,,
George F.R. Ellis*\,, and
William M. Lesame\dag\dag
}

\address{~}

\address{\dag\,Department of Fundamental Physics, University of
Barcelona, E-08028~Barcelona, Spain}

\address{\S\,Institute for Theoretical Physics, FSU Jena,
Max-Wien-Platz 1, D-07743~Jena, Germany}

\address{\ddag\,School of Computer Science and Mathematics, Portsmouth
University, Portsmouth~PO1~2EG, England}

\address{*\,Department of Mathematics and Applied Mathematics,
University of Cape Town, Cape Town~7701, South Africa}

\address{\dag\dag\,Department of Mathematics,
Applied Mathematics and Astronomy,
University of South Africa, Pretoria~0001, South Africa}

\address{~}

\date{15 April 1999}

\maketitle

\begin{abstract}

In a cold matter universe, the linearized gravito-magnetic tensor
field
satisfies a transverse condition (vanishing divergence)
when it is purely radiative. We show that in the
nonlinear theory, it is no longer possible to maintain the
transverse condition, since it leads to a non-terminating
chain of integrability conditions. These conditions are highly
restrictive,
and are likely to hold only in models with special symmetries, such as
the known Bianchi and $G_2$ examples.  In models with realistic
inhomogeneity, the gravito-magnetic field
is necessarily non-transverse at second and higher order.
\end{abstract}


\section{Introduction}

Gravitational waves in cosmology are usually
described by transverse traceless tensor perturbations of the
background Friedmann-Lema\^{\i}tre-Robertson-Walker (FLRW)
metric \cite{l}, i.e.
\[
ds^2=a^2(\eta)\left[-d\eta^2+\left(\gamma_{ij}+2{\cal H}_{ij}\right)
dx^idx^j\right]\,,
\]
where $a^2\gamma_{ij}$ is the background spatial metric,
$\gamma^{ij}{\cal H}_{ij}=0={\cal D}^j{\cal H}_{ij}$, and ${\cal D}$
is the covariant derivative formed from $\gamma_{ij}$.
An alternative covariant approach due to Hawking \cite{h}
is to describe gravitational radiation via the
electric and magnetic parts of the  Weyl tensor, which describes the
locally free part of the gravitational field.
The advantages of this approach are mainly its physical transparency,
and the readiness with which one can investigate nonlinear
extensions of perturbative theory. In particular, using the covariant
approach, one can investigate whether the various characteristic
properties of gravitational waves in the perturbative regime carry
through to the nonperturbative level.
This question is of importance for a covariant understanding
of second-order effects in cosmology.

In general terms, we expect that many linearized features will not
extend to
second order, precisely because of the nonlinearity of
the gravitational field, which ``creates its own background
on which it propagates". One consequence is that the separation of
inhomogeneous deviations into scalar, vector and tensor modes is not
in general consistent beyond the linearized level, and ``mode mixing"
occurs \cite{mmb}.  Indeed, in the fully nonlinear regime,
it is probably not possible to consistently isolate
the radiative part of the gravitational field.
Here we investigate whether the transverse property can be maintained
at second order   -- and we find that the answer is negative.
We show that  {\em in general the gravito-magnetic field in an
irrotational dust universe cannot be transverse in the exact nonlinear
theory.}  This implies in particular that {\em at second order,
the gravito-magnetic field  in cosmology is not in general
transverse.}
Thus there is unavoidable ``contamination" of tensor perturbations
with vector and scalar contributions via the divergence.

\section{Covariant approach to cosmological gravitational radiation}

The locally free gravitational field, i.e. that part of the
spacetime curvature which is not directly determined locally by
the energy-momentum tensor, is given by the Weyl tensor
$C_{abcd}$, which covariantly describes gravitational radiation
and tidal forces (see e.g. \cite{h,e,e1,mes}).
If $u^a$ is the cosmic average 4-velocity (with $u^au_a=-1$),
then comoving observers define the gravito-electric
and gravito-magnetic field tensors
\[
E_{ab}=C_{acbd}u^cu^d=E_{\langle ab\rangle}\,,~~
H_{ab}={\ts{1\over2}}\ep_{acd}C^{cd}{}{}_{be}u^e
=H_{\langle ab\rangle}\,,
\]
which exemplify a remarkable covariant electromagnetic analogy
in general relativity \cite{h,e1,mb}.
Here
$\ep_{abc}=\eta_{abcd}u^d$ is the spatial alternating tensor
formed from the spacetime alternating tensor $\eta_{abcd}$,
and the angled brackets on indices denote the
spatially projected, symmetric and
tracefree part:
\[
S_{\langle ab\rangle}= \left[h_{(a}{}^ch_{b)}{}^d-{\ts{1\over3}}
h^{cd}h_{ab}\right]S_{cd}\,,
\]
with $h_{ab}=g_{ab}+u_au_b$ the
spatial projector into the comoving rest space,
and $g_{ab}$ the spacetime metric.

In linearized theory, the gravito-electromagnetic fields give a
covariant
description of  the gravitational wave background in cosmology
\cite{h,dbe,he}, alternative to the
non-covariant description in standard perturbation
theory \cite{l,w,b}.
A necessary condition for these fields to describe radiation is that
their spatial curls should be nonzero; when the field is a pure
radiation field, additionally their spatial divergences should vanish,
in line with analogous properties of electromagnetic radiation
\cite{h,mb}. The covariant spatial divergence
and curl of tensors are defined by \cite{m}
\[
(\div S)_a=\D^bS_{ab}\,,~~\c S_{ab}=\ep_{cd(a}\D^cS_{b)}{}^d\,,
\]
where the
spatial part $\D_a$ of the covariant
derivative $\nabla_a$ is given by
\[
\D_a A_{b\cdots}=h_a{}^ch_b{}^d\cdots \nabla_c
A_{d\cdots} \,.
\]

Wave-propagation requires non-vanishing curls
\cite{dbe,he}, i.e.,
\begin{equation}
\c H_{ab} \neq 0 \neq \c E_{ab} \,,
\label{t'}
\end{equation}
if there is gravitational radiation present in the spacetime, allowing
the linking of gravito-electric and gravito-magnetic
variations to each other through
the Maxwell-like equations for these quantities, given below.
(More precisely, one requires $\c\c H_{ab}\neq0\neq\c\c E_{ab}$.
In \cite{mes}, it is shown that a further necessary condition is
nonvanishing distortions, i.e. $\D_{\langle a}H_{bc\rangle}\neq0
\neq\D_{\langle a}E_{bc\rangle}$. We will not need these refinements
here.)
In addition, if the gravitational radiation is ``pure",
i.e., if this radiation
encompasses the only deviation of the spacetime (at linear level)
from FLRW, then the covariant transverse conditions hold:
\begin{equation}
(\div H)_a=0=(\div E)_a \,.
\label{t}
\end{equation}
This is the simple, transparent and consistent
description of perturbative
gravitational radiation which flows from the covariant
approach \cite{h,dbe,he}.

In the nonperturbative theory,
we can impose the covariant conditions, but
we cannot a priori expect them to encode the same simple physical
meaning. If we can show that any one of the
conditions does not in general hold in the nonlinear case,
then since these conditions are necessary for pure radiation,
that is enough to show a breakdown in the linearized description
of gravitational radiation at nonperturbative level. Here we focus on
the gravito-magnetic transverse condition $(\div H)_a=0$, and
we show that it is in general inconsistent in the exact nonlinear
theory. We do not therefore need to investigate the gravito-electric
transverse condition $(\div E)_a=0$, although all the indications are
that this condition also breaks down at nonperturbative level.

We consider irrotational dust spacetimes, which provide
reasonable models of the late universe,
and allow us to focus on the purely gravitational dynamics of matter.
The known exact solutions with $(\div H)_a=0\neq \c H_{ab}$
are spatially homogeneous Bianchi type~V, as shown in  \cite{m},
and $G_2$ solutions with restricted inhomogeneity, found
in \cite{sv}. In \cite{mle1}, it was argued that in fact
$(\div H)_a=0$ does not impose any integrability conditions.
Although the basic equations and results of \cite{mle1} are
correct, there
is a flaw in the concluding argument, as we explain below.
A simple integrability condition implied by $(\div H)_a = 0$
  was assumed to be satisfied, but it is not identically satisfied
  in view of other valid equations. Rather
  it imposes non-trivial conditions which have to be checked.
Hence, as shown below,
the transverse condition cannot be obeyed in general,
although it can be at the linear level.

\section{Integrability of the gravito-magnetic transverse condition}

The correct statement is that in an irrotational dust universe,
$(\div H)_a=0$ implies the covariant integrability condition
\begin{equation}
{\cal I}^{(1)}\equiv [\sigma,\c H]=0 \,,
\label{1}\end{equation}
where $\sigma_{ab}=\D_{\langle a}u_{b\rangle}$
is the shear,
and $[A,B]$
denotes either the tensor commutator of spatial symmetric
tensors, i.e. $[A,B]_{ab}\equiv A_{ac}B^c{}_b-B_{ac}A^c{}_b$,
or its equivalent spatial dual:
\[
[A,B]_a
\equiv \ep_{abc}A^b{}_dB^{cd}={\ts{1\over2}}\ep_{abc}[A,B]^{bc} \,.
\]

To derive Eq. (\ref{1}) and its consequences, we need
some of the covariant
propagation and constraint equations, which are \cite{m}:
\begin{eqnarray}
\dot{\rho} &=& -\Theta\rho \,, \label{2}\\
\dot{\Theta} &=& -{\ts{1\over3}}\Theta^2-{\ts{1\over2}}\rho
-\sigma_{ab}\sigma^{ab} \,, \label{3}\\
\dot{\sigma}_{ab} &=& -{\ts{2\over3}}\Theta\sigma_{ab}-
\sigma_{c\langle a}\sigma_{b\rangle}{}^c-E_{ab} \,, \label{4}\\
\dot{E}_{ab} &=& -\Theta E_{ab}+3\sigma_{c\langle a}E_{b\rangle}{}^c
+\c H_{ab}-{\ts{1\over2}}\rho\sigma_{ab} \,, \label{5}\\
\dot{H}_{ab} &=& -\Theta H_{ab}+3\sigma_{c\langle a}H_{b\rangle}{}^c
-\c E_{ab} \,, \label{6}\\
(\div\sigma)_a &=& {\ts{2\over3}}\D_a\Theta \,, \label{7}\\
\c\sigma_{ab} &=& H_{ab} \,, \label{8}\\
(\div E)_a &=& [\sigma,H]_a+{\ts{1\over3}}\D_a\rho \,, \label{9}\\
(\div H)_a &=& -[\sigma,E]_a \,, \label{10}
\end{eqnarray}
where
$\rho$ is the energy density,
$\Theta=\D^au_a$ is the expansion rate, and
a dot denotes the covariant
time derivative $u^a\nabla_a$. (Note that
$\dot{S}_{ab}=\dot{S}_{\langle ab\rangle}$ since
the 4-acceleration vanishes.)

The general class of irrotational dust
spacetimes is characterized by
\[
p=0\,,~~\omega_a=0=q_a\,,~~\pi_{ab}=0\,,
\]
where $p$ is the isotropic
pressure, $\omega_a$ is the
vorticity, $q_a$ is the
energy flux, and $\pi_{ab}$ is the
anisotropic stress. These conditions do not lead to any
integrability conditions under time evolution,
as shown in \cite{m} (see also \cite{v,mac,ee,m2}).
By contrast, the class of shear-free dust spacetimes, characterized by
\[
p=0\,,~~q_a=0\,,~~\sigma_{ab}=0=\pi_{ab}\,,
\]
is not in general consistent. Time evolution of the
shear-free condition leads to the integrability condition
\[
\Theta\omega_a=0\,,
\]
as shown in \cite{e2} (see also \cite{sz,sss}).
More severe conditions arise from imposing
restrictions on the gravito-electric/magnetic fields.

In linearized theory,
$(\div H)_a$ vanishes
by virtue of Eq. (\ref{10}),
while $(\div E)_a=
{1\over3}\D_a\rho$,
i.e. density inhomogeneity is a source for the ``Coulomb"
part of the gravitational field. In order to isolate the
purely radiative
part in linearized theory, one sets the scalar inhomogeneity
$\D_a\rho$ to zero.
(Note that this does not imply there is no inhomogeneity in
the matter, only that the inhomogeneity is second order.)
Furthermore, in order to remove the remaining scalar and vector
modes, one sets $\D_a\Theta=0$, which by the
constraint equation (\ref{7}) requires $(\div\sigma)_a=0$.
It can then be shown that all these conditions corresponding
to purely tensor (radiative) modes do not produce
any integrability conditions, i.e. the covariant description
is consistent at the linear level
\cite{dbe,he}.

At nonperturbative level, $(\div H)_a$ is no
longer identically zero.
The vanishing of $(\div H)_a$
is equivalent
by Eq. (\ref{10}) to the algebraic condition
\begin{equation}
{\cal I}^{(0)}\equiv [\sigma,E]=0 \,,
\label{cond}\end{equation}
which is of course identically satisfied
at the linear level.
This covariant commutation property
expresses the fact that one can
choose
an orthonormal tetrad which is a simultaneous eigenframe
of $\sigma_{ab}$ and $E_{ab}$.
 Using the shear propagation
equation (\ref{4}) to eliminate $E$ in condition (\ref{cond}), we
find that
\[
[\sigma,\dot\sigma]=0\,,
\]
which allows one to further specialize the eigenframe to be
Fermi-Walker transported \cite{mle1}.
Conversely, if the shear eigenframe is Fermi-propagated, then
Eq. (\ref{cond}) necessarily follows, i.e. the gravito-magnetic
field is transverse \cite{mle1}.

It is possible to investigate consistency of the transverse condition
(\ref{cond}) using the special tetrad
\cite{s2},
but this leads to the same conclusion, with no less effort,
so we prefer to perform
a fully covariant analysis in which the physical and geometric
quantities themselves are to the fore.
The fundamental condition (\ref{cond}) must hold under
evolution along $u^a$. The first time derivative gives
\begin{equation}
\dot{\cal I}^{(0)}_a=-{\ts{5\over3}}\Theta{\cal I}^{(0)}_a-
{\ts{1\over2}}\sigma_a{}^b{\cal I}^{(0)}_b+{\cal I}^{(1)}_a\,,
\label{i1}\end{equation}
where
${\cal I}^{(1)}$ is defined in Eq. (\ref{1}).
To derive equation (\ref{i1}), we commute
the
shear propagation equation (\ref{4}) with $E_{ab}$,
and the gravito-electric
propagation equation (\ref{5}) with $\sigma_{ab}$,
using
basic algebraic properties of the commutator.
It follows from equation (\ref{i1}) that
the first time derivative of ${\cal I}^{(0)}$
is not automatically zero by virtue of the original
condition (\ref{cond}), but
yields the integrability condition (\ref{1}). Furthermore,
this integrability condition does not follow as a consequence
of the constraint equations.

Now the time derivative of the primary integrability condition
(\ref{1}) must also be satisfied. Evaluating this derivative
requires commuting $\c H_{ab}$ with the shear propagation equation
(\ref{4}), and $\sigma_{ab}$ with $(\c H_{ab})^{\rd}$.
There is no basic propagation equation for $\c H_{ab}$. However,
the propagation of $\c H_{ab}$ is indirectly determined by the curl of
the gravito-magnetic propagation equation (\ref{6}), together
with the identity \cite{m}
\begin{equation}
\left(\c S_{ab}\right)^{\rd}=\c \dot{S}_{ab}-{\ts{1\over3}}\Theta
\c S_{ab}-\sigma_e{}^c\ep_{cd(a}\D^eS_{b)}{}^d+3H_{c\langle a}
S_{b\rangle}{}^c \,.
\label{curl}\end{equation}
Then using Eqs. (\ref{6}) and (\ref{curl}), we find that
\begin{equation}
\dot{\cal I}^{(1)}_a=-2\Theta{\cal I}^{(1)}_a-\sigma_a{}^b
{\cal I}^{(1)}_b-{\cal I}^{(2)}_a\,,
\label{i2}\end{equation}
where
\begin{eqnarray}
{\cal I}^{(2)}_a &\equiv &[\sigma,\c\c E]_a
-3[\sigma,\c\langle\sigma H\rangle]_a+[E,\c H]_a+3H_a{}^b[\sigma,H]_b
\nonumber\\
{}&&{}+\sigma_{bc}H^{bc}\D_a\Theta-\left(H_{ac}\sigma^{cb}+
{\ts{1\over2}}\sigma_{ac}H^{cb}\right)\D_b\Theta
-\sigma_b{}^d\sigma^{bc}\D_cH_{ad}
\nonumber\\
{}&&{}+\left(\sigma_a{}^c\sigma_{bd}
+{\ts{1\over2}}\sigma_{ab}\sigma^c{}_d\right)\D_cH^{bd}\,.
\label{i3}\end{eqnarray}
Here $\langle AB\rangle_{ab}\equiv (AB)_{\langle ab\rangle}=
A_{c\langle a}B_{b\rangle}{}^c$, and we used the properties
$\ep_{abc}\ep^{def}=3!h_{[a}{}^d h_b{}^e h_{c]}{}^f$
and $\ep_{abc}\ep^{dec}=2h_{[a}{}^dh_{b]}{}^e$.
We can rewrite this expression
for ${\cal I}^{(2)}$ by using the identity \cite{m}
\begin{eqnarray*}
\c\c S_{ab} &=& -\D^c\D_cS_{ab}+
{\ts{3\over2}}D_{\langle a}(\div S)_{b\rangle}+
\left(\rho-{\ts{1\over3}}\Theta^2\right)S_{ab} \\
&&{}+3S_{c\langle a}\left\{E_{b\rangle}{}^c-{\ts{1\over3}}\Theta
\sigma_{b\rangle}{}^c\right\}+\sigma_{cd}S^{cd}\sigma_{ab}
-S^{cd}\sigma_{ca}\sigma_{bd}+\sigma^{cd}\sigma_{c(a}S_{b)d}\,,
\end{eqnarray*}
and the identity \cite{mes}
\[
\D_cS_{ab}=\D_{\langle c}S_{ab\rangle}-
{\ts{2\over3}}\ep_{dc(a}\c S_{b)}{}^d+{\ts{3\over5}}
(\div S)_{\langle a}h_{b\rangle c}\,.
\]
However the expression remains intractable.

Clearly satisfying the primary integrability condition (\ref{1})
on an initial surface
does not in itself guarantee that ${\cal I}^{(1)}$ remains zero.
Instead, the vanishing of $\dot{\cal I}^{(1)}$ requires,
by Eq. (\ref{i2}), an independent secondary integrability condition
\begin{equation}
{\cal I}^{(2)}=0 \,,
\label{i4}\end{equation}
which is of course identically satisfied at the linear level.
The form of Eq. (\ref{i3}) shows that the integrability conditions
grow more complicated at each stage, involving a growing
order of spatial derivative as well as increasing algebraic complexity.
The constraint equations do not lead to any simplification
of this integrability condition.

It is clear that further time evolution will produce a third, and then
a fourth, fifth, \dots integrability condition. Each condition is  more
complicated than its predecessor, and is not identically
satisfied in general by virtue of earlier conditions or the constraint
equations. Furthermore,
eliminating terms higher than second order does not change this
fundamental feature, although it does simplify the
integrability conditions somewhat. At second covariant order, the
primary integrability condition (\ref{1}) is unchanged in form,
but the secondary integrability condition (\ref{i4}) becomes
\begin{equation}
{\cal I}^{(2)}=[\sigma,\c\c E]+[E,\c H]=0\,.
\label{i4'}\end{equation}
The next integrability condition will involve third order spatial
derivatives, and so on.

\section{Conclusion}

We have shown that
the gravito-magnetic transverse  condition $(\div H)_a=0$,
taken over from linearized theory into the nonlinear theory, leads to
a chain of derived integrability conditions, the first two
of which are Eqs. (\ref{1}) and (\ref{i4}).
  This result corrects the assertion in \cite{mle1} that condition
  (\ref{1}) (and thus its derivatives) is identically satisfied.
  The logical flaw in the argument presented in
\cite{mle1} is essentially the implicit assumption that
the propagation equations (\ref{4})--(\ref{6}) are identically
satisfied. We have corrected that error here, by imposing the
propagation equations as effective conditions on the evolution
of $(\div H)_a=0$, equivalently $[\sigma,E]=0$.

The consequence of this is that {\em in general, the
gravito-magnetic field cannot be transverse at second (and higher)
order,
so that it cannot be purely radiative beyond linear order.}
In the fully nonlinear case, when one cannot really
isolate gravitational radiation, our result implies that
the transverse condition is in general inconsistent, i.e.
{\em generic inhomogeneity within the class of irrotational
dust spacetimes is inconsistent with a transverse gravito-magnetic
field.}
These results are independent of gauge choices or of coordinates,
since they are fully covariant.

There are however 
special cases when the integrability conditions
are satisfied, the most important of which is the case of almost-FLRW
universes. The condition (\ref{1}) and all its derivatives are
satisfied to first order in this case,
which simply reflects the fact that
the covariant formulation of the transverse condition
for perturbative gravitational
radiation is consistent at the linear level.

The integrability conditions are also satisfied if $H_{ab}=0$,
which defines the ``silent" models.  However, in that case, there
is another chain of integrability  conditions flowing from the
vanishing
of $H_{ab}$ at all derivative levels. As shown in \cite{eulem,s},
these integrability conditions are even more severe than in the
$(\div H)_a=0$ case. The $H_{ab}=0$
models are thus in general inconsistent (see also \cite{kp}),
and unlikely to extend beyond the known special cases where the
integrability conditions collapse to identities. These special
cases are some Bianchi and Szekeres solutions (see
\cite{eulem,s}).
Such a chain of derived integrability conditions of growing complexity
exists also in the
``anti-Newtonian" models $E_{ab}=0$ (see \cite{mle2}), where
the only known (irrotational dust) solution is the degenerate
FLRW solution. It is clearly very restrictive to impose conditions
on the gravito-magnetic and -electric fields of irrotational dust
spacetimes.

In the $(\div H)_a=0\neq H_{ab}$ case, among the
known special exact solutions are Bianchi type~V \cite{m} and
diagonal (in comoving coordinates) $G_2$
solutions \cite{sv}. In these solutions, the integrability
conditions are identically satisfied by virtue of the special
structure of $\c H_{ab}$. If we choose a Fermi-propagated
shear eigenframe (as described above), then in these solutions,
$\c H_{ab}$ remains diagonal at all times, so that the
primary integrability condition (\ref{1}) is identically
true, as are all its derivatives. For example, for $G_2$ solutions
of the form
\[
ds^2=-dt^2+A^2(t,x)dx^2+B^2(t,x)dy^2+C^2(t,x)dz^2\,,
\]
in coordinates that are comoving with the fluid 4-velocity $u^a$,
it follows that
\[
H_{ab}=-2C\left({B_{,x}\over A}\right)_{\!,t}
\delta_{(a}{}^y\delta_{b)}{}^z\,,
\]
which leads to a diagonal $\c H_{ab}$.
The Bianchi type~V solution has $\c H_{ab}\propto\sigma_{ab}$
in the shear eigenframe. In the Bianchi case, it is
possible to identify further examples \cite{les}. The type~V case is
distinguished as the only $(\div H)_a=0\neq H_{ab}$ solution in
Bianchi class~B, and it has $(\div E)_a\neq0$,
so that it is not purely radiative in the sense of
condition (\ref{t}).
(Note that the diagonal $G_2$ solutions also have $(\div E)_a\neq0$.)

Indeed it is clear from our analysis that a similar chain
of integrability conditions will flow from the
gravito-electric transverse condition in Eq. (\ref{t}),
which is equivalent to
\[
 {\textstyle{1\over3}}\D_a\rho+[\sigma,H]_a=0 \,.
\]
The independent chain of conditions arising from repeated time
differentiation of this condition reinforces the unlikelihood
of finding inhomogeneous
transverse radiative solutions beyond the linear level.

The Bianchi class~A solutions all have vanishing
divergence for both $E_{ab}$ and $H_{ab}$, and so
provide spatially homogeneous pure radiative examples,
except for types~I and LRS~VII$_0$, which are in the
degenerate
silent case $H_{ab}=0$.
Thus {\it in Bianchi class~A spacetimes
it is possible to find exact solutions satisfying
both the radiative transverse conditions~(\ref{t})
and the propagating radiative conditions~(\ref{t'})}.
We give here two such examples, using the formalism and
notation of \cite{em}.
In a shear eigenframe $\{\bf{e}_\alpha\}$ (where $\alpha=1,2,3$),
a tracefree diagonal tensor $S_{\alpha\beta}$ has curl
\begin{equation}
\c S_{\alpha\beta} =
\ep_{\gamma\delta(\alpha}\partial^\delta S_{\beta)}
{}^\gamma+\case{1}/{2}S_{\alpha\beta}n^{\delta}{}_{\delta}
-3 n^{\gamma}{}_{(\alpha}S_{\beta)\gamma}
+\delta_{\alpha\beta}S_{\gamma\delta}n^{\gamma\delta} \,,
\label{bianchi}\end{equation}
where $\partial_\alpha$ is the directional derivative along
$\bf{e}_\alpha$, and
the tetrad commutation quantities may be
diagonalized: $n_{\alpha\beta}={\rm diag}(n_1,n_2,n_3)$.
Using the irreducible components
$S_{+} = -{3\over2}S_{11}$ and  $S_{-} =
{1\over2}\sqrt{3}(S_{22}-S_{33})$,
the Bianchi type~II solutions (characterized by
$n_{1}>0$ and $n_{2}=0=n_{3}$)
have nonvanishing components of $\c S_{\alpha\beta}$
\[
\c S_+ = -\case{3}/{2}n_{1}S_+ \,, ~~
\c S_- = \case{1}/{2}n_{1} S_-  \,,
\]
and nonvanishing components of $\c \c S_{\alpha\beta}$
\[
\c \c S_+ = \case{9}/{4}(n_1)^2 S_+ \,, ~~
\c \c S_- = \case{1}/{4}(n_1)^2 S_- \,.
\]
The type~VI$_{0}$ solutions with $n^{\alpha}{}_{\alpha}=0$
(and $n_2=-n_1<0$, $n_3=0$),
have nonvanishing components of $\c S_{\alpha\beta}$
\[
\c S_+ = -\case{1}/{2}\sqrt{3}n_1(\sqrt{3}S_{+} - S_-) \,, ~~
\c S_- = \case{1}/{2}\sqrt{3}n_1(S_+ +\sqrt{3} S_-) \,,
\]
and nonvanishing components of $\c \c S_{\alpha\beta}$
\[
\c \c S_\pm = 3(n_1)^2 S_\pm \,.
\]
The tensor $S_{\alpha\beta}$ in both examples
may then be chosen to be either the gravito-electric
tensor $E_{\alpha\beta}$ or gravito-magnetic
tensor $H_{\alpha\beta}$.

The importance of these examples is twofold: (a) they show that exact
purely radiative solutions can exist in the minimal covariant sense,
and (b) they show that the Petrov type~N characterization
of a radiative field
is more restrictive,
because there are no Petrov type~N dust metrics in the full
nonlinear theory \cite{kt,sz2}.
However, because these examples are spatially homogeneous models,
information is
not being conveyed from place to place by the waves; these are
``standing" waves rather than traveling (propagating) waves.
By contrast, in the linearized case, arbitrary information can be
conveyed by the gravitational waves \cite{he}.

It is possible that further, inhomogeneous exact solutions
can be found,
but they would also be very special cases,
in order to satisfy the chain of integrability conditions.
Realistic inhomogeneous models cannot be expected to
satisfy these conditions.\\

{\bf Acknowledgements:}
Thanks to Bruce Bassett, Henk van Elst, David Matravers
and Jose Senovilla for very useful comments.
CFS was supported by the Alexander von Humboldt Foundation.


\end{document}